\documentclass[conference]{IEEEtran}
\IEEEoverridecommandlockouts

\usepackage{cite}
\usepackage{amsmath,amssymb,amsfonts}
\usepackage{algorithmic}
\usepackage{graphicx}
\usepackage{hyperref}
\usepackage{textcomp}
\usepackage{xcolor}
\def\BibTeX{{\rm B\kern-.05em{\sc i\kern-.025em b}\kern-.08em
    T\kern-.1667em\lower.7ex\hbox{E}\kern-.125emX}}
\begin{document}

\title{\LARGE \bf
CAN-STRESS: A Real-World Multimodal Dataset for Understanding Cannabis Use, Stress, and Physiological Responses\thanks{This work was supported in part by the National Science Foundation under grant CNS-2210133. Funding for the data collection study was provided by the Alcohol and Drug Research Program (ADARP) of Washington State University. Any opinions, findings, conclusions, or recommendations expressed in this material are those of the authors and do not necessarily reflect the views of the funding organization.}
}

\author{Reza Rahimi Azghan$^{1}$, Nicholas C. Glodosky$^{2}$, Ramesh Kumar Sah$^{2}$, Carrie Cuttler$^{2}$, \\ Ryan McLaughlin$^{2}$, Michael J. Cleveland$^{2}$, Hassan Ghasemzadeh$^{1}$
\thanks{$^{1}$Arizona State University, Phoenix, AZ, USA}%
\thanks{$^{2}$Washington State University, Pullman, WA, USA}%
\thanks{{email: \{rrahimia, hassan.ghasemzadeh\}@asu.edu, \{nicholas.glodosky, ramesh.sah, carrie.cuttler, ryan.mclaughlin, michael.cleveland\}@wsu.edu}}%
}

\maketitle

\begin{abstract}

Coping with stress is one of the most frequently cited reasons for chronic cannabis use. Therefore, it is hypothesized that cannabis users exhibit distinct physiological stress responses compared to non-users, and that these differences may be especially pronounced during moments of cannabis consumption. However, there is a scarcity of publicly available datasets that allow such hypotheses to be tested under real-world conditions. This paper introduces a dataset named \textit{CAN-STRESS}, collected using Empatica E4 wristbands. The dataset includes multimodal physiological measurements (such as skin conductance, heart rate, and skin temperature) from 82 participants (39 cannabis users and 43 non-users) as they went about their daily routines. In addition to sensor data, participants provided self-reported survey responses that included perceived stress ratings and timestamps of key daily events such as cannabis use, physical activity, and sleep. To demonstrate the utility of the dataset for downstream applications, we present a preliminary machine learning task aimed at classifying cannabis users versus non-users based on physiological features. Our model achieves a classification accuracy of approximately 96\% and an f1-score of around 98\%. An analysis of feature importance using SHAP values revealed that electrodermal activity and heart rate metrics were the most influential predictors, consistent with their established roles in stress detection. We publicly release the CAN-STRESS dataset, which we believe serves as a reliable and rich resource for studying the physiological correlates of cannabis use and stress in naturalistic settings.

\end{abstract}

\begin{IEEEkeywords}
Physiological Data, Wearables, Stress, Machine Learning
\end{IEEEkeywords}
\section{Introduction}
Due to its widespread consumption and the increasing legalization and decriminalization in many regions, studying cannabis use and its implications is essential \cite{xiao2023cannabis}. Understanding the physiological and psychological effects of cannabis is crucial for informing policy, healthcare, and individual decision-making. Furthermore, differentiating between the physiological responses of cannabis users and non-users offers insights into how chronic use may alter stress regulation, a frequently cited reason for cannabis consumption \cite{al2021impact}. However, most existing studies are conducted in controlled laboratory environments, where ecological validity is constrained \cite{meier2024cannabis}. These settings fail to capture the nuanced, real-world conditions under which individuals consume cannabis, including the interplay of stressors and other contextual factors. Bridging this gap through field-based research is imperative to develop a comprehensive understanding of cannabis effects under realistic settings and enhance the applicability of findings to broader populations and real-world scenarios.

\begin{figure}
    \centering
    \includegraphics[width=1\linewidth]{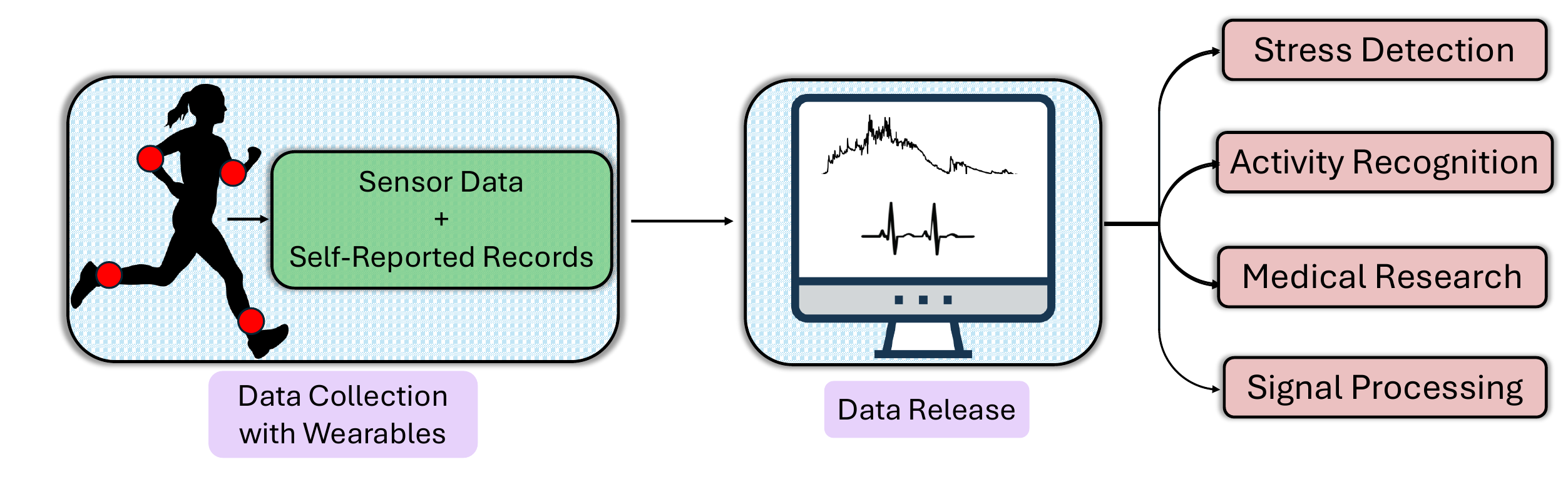}
    \caption{The process of collecting physiological data using wearable devices, storing it in a centralized web portal, and enabling access for various research applications addresses the limitations of laboratory-based studies. The stored data can be used to advance research in fields such as stress detection, activity recognition, and medical diagnostics.}
    \label{fig:no-labels}
\vspace{-7mm}
\end{figure}

The CAN-STRESS dataset was developed to enable the study of cannabis use and stress in ecologically valid, real-world conditions. It consists of multimodal data from 82 participants who wore Empatica E4 wristbands~\cite{empaticae4} for 24 hours while engaging in daily activities. The dataset includes electrodermal activity (EDA), heart rate (HR), body temperature, and accelerometer data, synchronized with self-reported logs of cannabis use, sleep, exercise, and perceived stress levels. By including both frequent cannabis users and non-users, CAN-STRESS provides a balanced framework for comparative analyses and is one of the largest publicly available datasets capturing the physiological characteristics of cannabis users in natural environments.

Participants were recruited with strict inclusion and exclusion criteria. Cannabis users reported daily or near-daily use ($\geq$4 times per week for at least one year), while non-users reported fewer than 10 lifetime uses and none in the past year. Individuals were excluded if they had neurological disorders, psychosis, autism, bipolar I, heavy alcohol consumption, recent use of illicit drugs, nicotine, or corticosteroid medications. Eligible participants were at least 21 years old, fluent in English, and smartphone owners. The study was reviewed and approved by the Washington State University Institutional Review Board, and informed consent was obtained prior to participation. To protect participant privacy, all wearable and survey data were anonymized and securely stored, with study materials collected directly by researchers to minimize risk of exposure.

Several studies have already leveraged CAN-STRESS to advance cannabis research. ~\cite{mybsn} demonstrated that EDA signals can be used to detect cannabis consumption episodes in naturalistic environments. ~\cite{azghan2024cudle} introduced the CUDLE framework, showing that self-supervised learning can efficiently detect cannabis use with limited labeled data. A third work examined stress regulation, revealing disrupted diurnal stress rhythms among cannabis users and showing that cannabis decreased stress in daily life, in contrast to laboratory findings~\cite{glodosky2024multimodal}. Together, these works highlight the dataset’s value in enabling ecologically valid and computationally innovative research.

Building on this foundation, the present paper contributes (1) a detailed description of the dataset, (2) new analyses comparing physiological features of users vs. non-users, and (3) a baseline machine learning pipeline that distinguishes users from non-users using individual-level features. These contributions aim to broaden access to the dataset and establish benchmarks for future computational health research.
\section{Data Modalities}

\begin{figure*}
    \centering
    \includegraphics[width=0.8\linewidth]{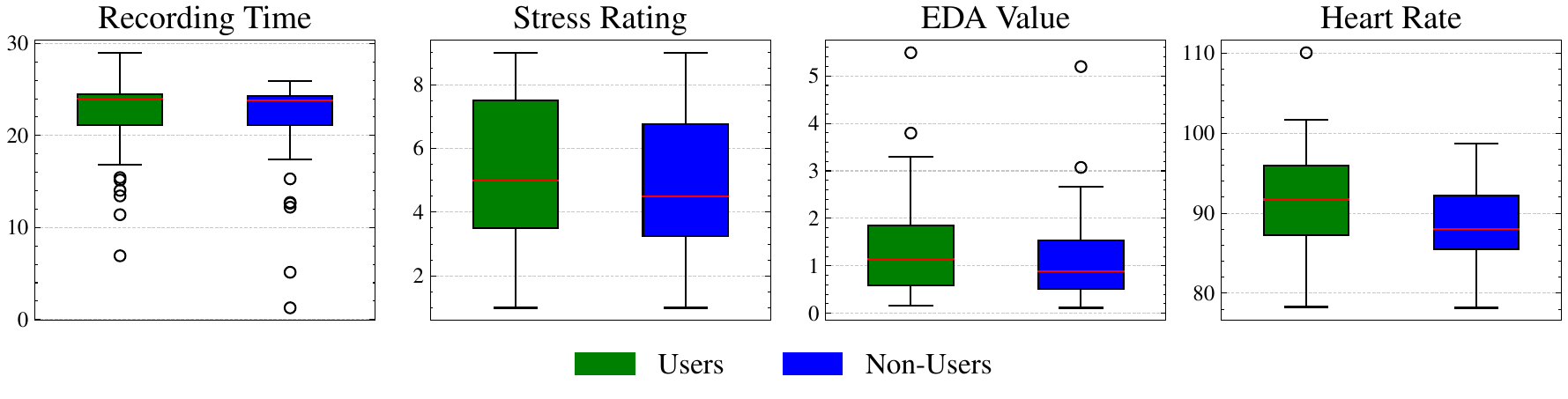}
    \caption{Comparison of dataset features between cannabis users (n=39) and non-users (n=43). Each boxplot represents participant-level summary values: (1) total recording duration (hours), (2) mean self-reported stress rating across the day, (3) mean electrodermal activity (EDA, µS), and (4) mean heart rate (bpm)}
    \label{fig:features_boxplot}
\vspace{-4mm}
\end{figure*}

The CAN-STRESS dataset includes two primary modalities: a self-reported questionnaire and multimodal physiological data collected through the E4 wearable wristband. Together, these modalities provide a comprehensive view of participants' activities, physiological responses, and subjective experiences. Aligning and integrating the two sources of the dataset allows researchers to identify and analyze the correlations between different events, e.g., cannabis consumption and exercise.

 The dataset is organized in a straightforward structure to facilitate use by other researchers. All files are arranged under a root directory labeled \texttt{CAN-STRESS/}, with a subfolder for each anonymized participant identifier. Within each participant’s folder, the physiological modalities are stored as individual \texttt{.csv} files (e.g., \texttt{ACC.csv}, \texttt{EDA.csv}, \texttt{BVP.csv}), each containing time-stamped recordings sampled at their respective rates. To aid interpretation, each participant folder also includes an \texttt{info.txt} file describing the contents, units, and sampling rates of the CSV files. The self-reported questionnaire data are compiled across all participants in a single \texttt{logbook.xlsx} file stored at the root level, which documents daily activities such as cannabis use, sleep, exercise, and stress ratings. 

\subsection{Self-Reported Questionnaire}
The first modality consists of a structured questionnaire that participants completed during the data collection period. These self-reported entries provide timestamps and labels that can be matched with the second modality to facilitate the analysis of relationships between physiological signals and daily activities. The questionnaire captures significant moments throughout the participants' day, including:

\begin{itemize}
\item \textbf{Sleep Patterns}: Participants recorded the times they went to bed and woke up, which provided insight into their sleep duration and routines.
\item \textbf{Cannabis Use}: Participants stated whether or not they were a user, and they also documented the start and end times of their cannabis consumption.
\item \textbf{Exercise Activities}: The peak moments of physical activity during the day were also recorded, which in turn enabled us to account for physiological changes associated with these activities. 

\item \textbf{Stress Ratings}: At key moments (e.g., sleep, exercise, cannabis use), participants rated their perceived stress on a scale from 1 (not at all stressed) to 10 (extremely stressed). This straightforward approach has been widely used in ecological momentary assessment (EMA) studies to capture in-the-moment subjective stress levels~\cite{shiffman2008ecological}.

\end{itemize}

\subsection{Physiological Data from the Wristband}
The second modality includes continuous physiological data collected using the Empatica E4 wristband, a medical-grade wearable device. This modality consists of the following signals:
\begin{itemize}
\item \textbf{Accelerometer (ACC)}: Captures triaxial movement data at 32 Hz, which can be used to detect activity patterns, such as exercise or sedentary behavior.
\item \textbf{Blood Volume Pulse (BVP)}: Measured at 64 Hz, BVP provides raw data for calculating heart rate and interbeat intervals that offers insights into cardiovascular dynamics.
\item \textbf{Electrodermal Activity (EDA)}: Collected at 4 Hz, EDA measures skin conductance, which is closely associated with stress and emotional arousal \cite{pop2020electrodermal}.
\item \textbf{Heart Rate (HR)}: Derived from BVP, HR is sampled at 1 Hz and reflects real-time changes in cardiovascular activity.
\item \textbf{Interbeat Interval (IBI)}: Derived from BVP, IBI represents the time between successive heartbeats and is critical for heart rate variability analysis.
\item \textbf{Body Temperature (TEMP)}: Recorded at 4 Hz, TEMP tracks changes in skin temperature, which may be indicative of physiological or environmental changes.
\end{itemize}

These physiological signals provide high-resolution, multivariate data that capture participants’ physical states and responses in real time. By combining the self-reported data with wristband signals, researchers can investigate the interplay between subjective experiences (e.g., stress ratings) and objective physiological responses (e.g., changes in HR or EDA) across various activities and contexts.
\section{Descriptive Analysis of Cannabis Users and Non-Users}

To provide a comprehensive overview of the dataset, we present key statistical features that highlight differences between cannabis users and non-users. These analyses help characterize both baseline physiological patterns and stress-related responses, and serve as a foundation for downstream computational tasks. In particular, we use these features to develop a machine learning model that classifies cannabis users and non-users based on their physiological data.

\subsection{Group-Level Trends}

This subsection presents important statistical distinctions between cannabis users and non-users, covering aspects such as average recording time, self-reported stress levels, sleep duration, and physiological measurements including EDA and heart rate. These variables reflect common behavioral and physiological characteristics within each group and form the basis for additional analyses, including those involving predictive modeling.

Figure~\ref{fig:features_boxplot} presents four boxplots comparing cannabis users (n=39) and non-users (n=43) across key physiological and behavioral dimensions. Recording duration (in hours) reflects the total amount of usable data collected per participant. For all other measures, participant-level means were computed over the 24-hour recording period, including average self-reported stress ratings (on a 1–10 scale), mean electrodermal activity (EDA, in microsiemens), and mean heart rate (in beats per minute). These participant-level summary values were then aggregated within each group and visualized as boxplots to illustrate variability and group-level trends. The results highlight consistent patterns, with cannabis users reporting higher stress ratings and exhibiting elevated EDA and heart rate values compared to non-users.

\subsection{Machine Learning for User Classification}

Building on the group-level feature analysis, we explore the use of machine learning to classify participants as cannabis users or non-users based on their physiological data. The goal of this task is not to develop a production-ready classifier but to demonstrate the feasibility of using wearable data for downstream predictive modeling. The envisioned pipeline takes as input the raw physiological signals collected by the wristband and predicts a participant’s user status. For personalization, we fine-tuned the pre-trained model using 50\% of the available windows from each test subject, while reserving the remaining 50\% for evaluation. This setup highlights the potential of the dataset for real-world applications that rely on minimal supervision and passive sensing. 

The classification model employed a multi-layer perceptron architecture consisting of three hidden layers with 128, 64, and 32 neurons, respectively, followed by a single-neuron output layer for binary classification. Each hidden layer incorporated batch normalization and ReLU activation, with dropout regularization (rate=0.2) applied to mitigate overfitting. The output layer utilized sigmoid activation to produce probability scores. Training and fine-tuning were performed using the Adam optimizer, binary cross-entropy loss, and a learning rate of 0.001.

To prepare the input data for our model, we selected four physiological modalities from the wristband data for our predictive machine learning task: EDA, BVP, ACC, and TEMP. To ensure data relevance and minimize confounding from sleep-related physiological changes, we restricted our analysis to data recorded during participants' waking hours. The continuous signals were segmented using a sliding window approach with fifteen-minute windows and 50\% overlap between consecutive segments~\cite{stressalyzer}. On average, this produced approximately 180 windows per participant, resulting in a total of about 14,600 windows across the dataset. Each window was represented by 31 physiological features extracted from the EDA, BVP, ACC, and TEMP signals.

For feature extraction, we applied modality-specific methods designed to capture the distinct temporal and physiological characteristics of each signal. Using the NeuroKit2~\cite{Makowski2021neurokit} package, we decomposed the EDA signal into tonic and phasic components and extracted features such as mean tonic level, number of skin conductance response peaks, peaks per minute, and the statistical properties of peak amplitudes. From the BVP signal, we derived heart rate variability (HRV) features, including RMSSD, pNN50, and SDNN, in addition to basic statistical measures of heart rate (e.g., mean and standard deviation). For the temperature signal, we computed statistical metrics, including mean, standard deviation, and range, along with the linear slope to capture thermal trends across each window. Accelerometer data was processed by computing the vector magnitude of the triaxial acceleration, from which we extracted features such as mean activity level, proportion of time spent in motion, and axis-specific statistics that characterize both intensity and direction of movement. In total, we extracted 31 features across all modalities, which served as the input to our downstream classification task.

Following feature extraction, our dataset consisted of tabular data, where each row corresponded to a 15-minute window and included 31 physiological features. To address individual differences in baseline physiology, we applied subject-wise standardization, ensuring that each participant’s features had a mean of zero and a standard deviation of one.
We employed a leave-one-subject-out strategy to evaluate model generalization to new individuals. In each fold, one participant served as the test subject while the model trained on all remaining participants. To personalize the model for each test subject, we implemented a transfer learning approach: we first trained a base model on the training subjects, then fine-tuned the final layer of the pre-trained model using 50\% of the test subject's data, and evaluated performance on the remaining 50\%. This process was repeated across all participants, with each individual serving as the test subject once. Final performance metrics were computed as averages across all folds.

In table \ref{table:performance_comparison}, We report our model's performance on both the training and test data using four evaluation metrics: accuracy, F1 score , precision, and recall. Accuracy represents the overall proportion of correct predictions. Precision measures the fraction of predicted positive cases that are actually positive, while recall reflects the fraction of actual positive cases that were correctly identified by the model. The F1 score is the harmonic mean of precision and recall, offering a single metric that balances both, particularly useful in scenarios with class imbalance.

\begin{table}[h]
\caption{Model performance on training and test data (averaged across subjects).}
\centering
\small 
\resizebox{0.5\textwidth}{!}{%

\begin{tabular}{ccccc}
\hline
\textbf{Data} & \textbf{Acc} & \textbf{Prec} & \textbf{Rec} & \textbf{F1} \\
\hline
Train Data & 99.93\% ($\pm$0.00) & 99.96\% ($\pm$0.00) & 100.00\% ($\pm$0.00) & 99.93\% ($\pm$0.05) \\
Test  Data & 95.96\% ($\pm$0.06) & 97.82\% ($\pm$0.05) & 100.00\% ($\pm$0.00) & 95.92\% ($\pm$0.06) \\
\hline
\end{tabular}}
\label{table:performance_comparison}
\vspace{-3mm}
\end{table}

\begin{figure}
    \centering
    \includegraphics[width=\linewidth]{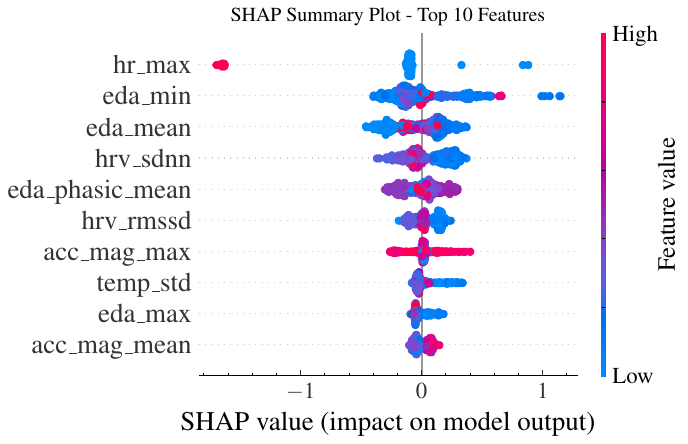}
\caption{SHAP summary plot showing the top 10 most important features influencing the model’s predictions. Features related to heart rate, heart rate variability, and electrodermal activity contribute most strongly}
    \label{fig:shap_summary}
\vspace{-5mm}
\end{figure}

To better understand which signals influenced the model’s predictions, we used SHAP (Shapley Additive Explanations)~\cite{lundberg2017unified} to analyze feature importance. As shown in the SHAP summary plot (Figure~\ref{fig:shap_summary}), features derived from HR and EDA dominated the top rankings. The most impactful feature was the maximum heart rate (\texttt{hr\_max}), followed closely by several EDA-related features, including \texttt{eda\_min}, \texttt{eda\_mean}, \texttt{eda\_phasic\_mean}, and \texttt{eda\_max}. Heart rate variability metrics such as \texttt{hrv\_sdnn} and \texttt{hrv\_rmssd} also contributed significantly. This result aligns with the expectation that physiological markers of stress (captured through both heart rate and skin conductance) differ between cannabis users and non-users. Overall, the SHAP results highlight the central role of stress-related signals, particularly EDA and HR, in driving the model’s ability to distinguish between the two groups.

\section{Conclusion}
In this paper, we introduced CAN-STRESS, a multimodal dataset designed to advance research into the physiological and behavioral effects of cannabis use in real-world settings\footnote{\href{https://zenodo.org/records/14842061}{https://zenodo.org/records/14842061}}. By integrating self-reported data on key daily activities with high-resolution physiological signals collected via a wearable wristband, CAN-STRESS provides a unique opportunity to examine the relationship between subjective experiences and objective measurements. Addressing the limitations of lab-based studies, CAN-STRESS offers an ecologically valid foundation for exploring stress regulation, activity recognition, and other health-related research domains. We encourage researchers to utilize CAN-STRESS to drive advancements in behavioral science, wearable computing, and medical diagnostics.

\bibliographystyle{IEEEtran}
\bibliography{refs}

\end{document}